\documentclass[prd,twocolumn,amsmath,amssymb]{revtex4}

\usepackage{graphicx}
\usepackage{dcolumn}
\usepackage{bm}

\begin{document}



\title{Non-converging hysteresis cycles in random spin networks \\ \vspace*{0.0cm}}

\author{Ondrej Hovorka}

\author{Gary Friedman}
\affiliation{Electrical and Computer Engineering Department, Drexel University, Philadelphia, PA 19104, USA}


\date{\today}
           
\begin{abstract}
Behavior of hysteretic trajectories for cyclical input is investigated as a function of the internal structure of a system modeled by the classical random network of binary spins. Different regimes of hysteretic behavior are discovered for different network connectivity and topology. Surprisingly, hysteretic trajectories which do not converge at all are observed. They are shown to be associated with the presence of specific topological elements in the network structure, particularly with the fully interconnected spin groups of size equal or greater than 4. \\




\end{abstract}


\maketitle

\newpage
Although hysteresis is often illustrated by closed loops, there is no fundamental reason to expect that periodically varying external parameters should produce any kind of closed hysteretic trajectories in general. In fact, hysteretic trajectories which do not converge are observed experimentally in many systems. It is known, for example, that repeated mechanical loading and unloading of some materials leads to gradual accumulation of strain and eventual material failure, rather than stable hysteresis cycles \cite{r1}. Similar non-convergent behavior in magnetic materials is manifested as a gradual shift of magnetization occurring under the action of periodic external field \cite{r2}. The question arises: What in the structure of a system determines the existence of non-convergent hysteretic trajectories?\\
\indent Here this issue is investigated by looking at binary spin networks as a way to model the structure of a system. Although it is most frequently employed to model magnetic materials, binary spin network is a classical prototype for studying hysteretic properties of many natural systems. Hysteretic cycles have been studied previously using the Random Field Ising model \cite{r3} and random networks of antiferromagnetically coupled spins \cite{r4} where the focus was mainly on the effects of disorder. In addition, interesting memory effects have been discovered in spin glass networks such as complementary point and reversal-field memories \cite{r5}. Particularly important property found to occur in ferromagnetic \cite{r6} and some antiferromagnetic networks \cite{r7} is the return point memory (RPM), which is responsible for the recovery of the microscopic state and formation of closed minor hysteresis loops after the very first external field cycle. Networks without RPM have been found to display only a gradual minor loop formation or sub-harmonic cycles \cite{r8} (cycles with period multiple of the external field period). Despite the substantial amount of work performed in this area, ability of binary networks to describe non-convergent hysteretic trajectories was not observed. The link between the network structure and the convergence character of hysteretic trajectories also remains largely unexplored.\\
\indent In this Letter, investigating the link between the network connectivity and topology and its hysteretic behavior, we discover growth of non-convergent trajectories with emergence of certain topological structures. Specifically we find that RPM observed at low network connectivity gives way to transient trajectories converging to sub-harmonic cycles or to trajectories with very long transient time growing without bound as the system size increases and thus becoming non-convergent.\\
\indent As a prototypical system, we consider classical Erd$\ddot{\rm o}$s-R\'enyi (ER) random network of anti-ferromagnetically coupled spins. It is one the simplest realizations of a complex network. It is also convenient because its connectivity and topological structure can be tuned by adjusting a single parameter - probability of connection between pairs of spins \cite{r9}. Generally, a random network of $N$ spins can be viewed as a random graph with $N$ nodes (vertices) connected by undirected edges representing pair-wise interactions between the spins. ER networks are constructed using a random graph process where edges are added one by one between randomly chosen pairs of spins. The structure of the graph is characterized by its adjacency matrix $A_{ij}$ with elements $A_{ij}=1$ if the vertices $i$ and $j$ are connected or $A_{ij}=0$ if the vertices are disconnected. Antiferromagnetic interactions between the spins will then be described by a matrix $J_{ij}=-JA_{ij}$, where the $J$ is a positive number representing the interaction strength.\\
\indent Dynamics of the system is governed by minimization of the standard Ising-type energy:
\begin{equation}
G=-\frac{1}{2}\sum_{ij}J_{ij}s_is_j-H\sum_is_i+\sum_if_i(s_i),
\label{eq:one}
\end{equation}
where $s_i$ denotes the $\pm 1$ spin variable and $H$ is the external field. The first and the second sum in~(\ref{eq:one}) are respectively the spin-spin and the spin-external field interaction energies. Additionally, there is a symmetric double-well potential energy $f_i$, due to which any isolated spin $s_i$ flips from $-1$ to $+1$ at the field-threshold $+\alpha_i$, while the flipping from $+1$ to $-1$ occurs at the field $-\alpha_i$. The positive threshold magnitudes $\alpha_i$ will be viewed here as random variables to mimic a structural disorder in the system. Note that the difference between the `up' and `down' thresholds, $+\alpha_i>-\alpha_i$, results in an inherent hysteretic behavior of each spin. Such classical hysteretic spins are frequently used as representations of single-domain magnetic grains, tiny capillary pores in absorbing materials, vortex pinning imperfections in superconductors, individual decision making agents in socio-economic systems, etc. They are also the main building blocks of some phenomenological models of hysteresis \cite{r10}.\\
\indent The local field $h_i$ governing the flipping of each spin is obtained as a variation of Eq.~(\ref{eq:one}) with respect to $s_i$ and equals:
\begin{equation}
h_i=-\delta G/\delta s_i=-J\sum_j A_{ij}s_j+H
\label{eq:two}.
\end{equation}
We consider standard field driven adiabatic dynamics \cite{r6} where, at each time step, the spin states are updated only if $h_is_i<-\alpha_i$. For a network with negative interactions such dynamics may produce avalanches with backward flipping spins when $\Delta H\Delta s_i<0$ resulting in a non-monotonic variation of state of the entire system even when the field $H$ varies monotonically \cite{r11}. Here we avoid such backward spin flips by restricting dynamic behavior to single spin flips. According to~(\ref{eq:two}) the spin back-flips do not occur as long as the inequality $J\cdot\max(d_i)<\min(\alpha_i)=\alpha_{\min}$ is satisfied for the particular network, where $d_i = \sum_jA_{ij}$ is a degree of vertex $i$ in the network, i.e. the coordination number of the spin $s_i$. Hence, choosing $J<J_t=\alpha_{\min}/\max(d_i)$ guarantees the absence of back-flips, resulting in a simple avalanche behavior consisting of single spin transitions $\Delta H\Delta s_i>0$ and monotonic state variation. In simulations, we use Gaussian distribution of $\alpha_i$ with variance $\sigma$ and mean $\mu >> \sigma$ \cite{r12}. Such an assumption is natural for various realistic systems e.g. magnetic films with very strong perpendicular anisotropy or patterned nanostructures \cite{r13}.\\
\indent It turns out that even the single spin-flip dynamics limit introduced above yields complex minor loop behavior. This is demonstrated here by simulating hysteretic trajectories corresponding to periodic cycles of the external field (Fig. 1) obtained by increasing the field starting from the negative saturation, where all spins are in the $-1$ state, to the point $H_r$ with the average spin state (magnetization) $\langle S\rangle$, then decreasing the field to $-H_r$ and returning back to $H_r$. The ability of the network to recover its state when the field returns to $H_r$ will be quantified by the cycle opening, denoted by $\Delta S$ and equal to \%-difference in a number of spins which did not return to the original state.
\begin{figure}
\includegraphics[scale=0.8]{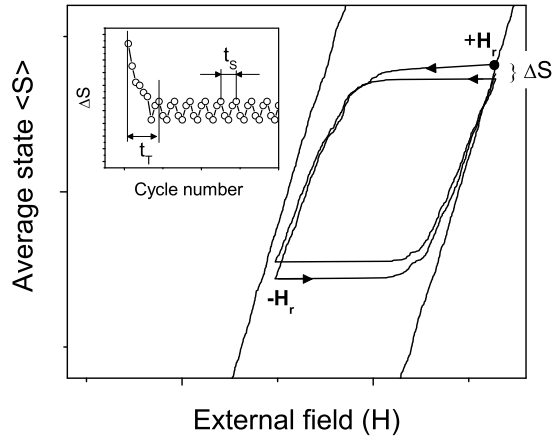}
\caption{\label{fig:epsart1} Hysteretic cycles for two periods of external field. $S$ and $\langle S\rangle$ denote, respectively, the spin microstate state and the average state (magnetization). $\Delta S$ is a measure of the cycle opening calculated by comparing spin patterns before and after each cycle. Results were obtained from simulation of $50^2$ random spin-network with the average coordination number (degree) $d=\langle d_i\rangle = 10$. Reducing the degree to $d<1$ results in the immediately closed minor loops $(\Delta S=0)$. Inset: Dependence of $\Delta S$ on the external field period number. The steady state reached after initial transient time $t_T$ can contain simple minor loops with $t_S=1$ or subharmonic cycles with $t_S >1$.}
\end{figure}\\
\indent We observe that the cycle opening $\Delta S$ depends on the degree $d$ (i.e. connectivity) of the network, which is defined as an average over all $d_i$ in the particular network realization. For connectivity $d << 1$, the system returns to the same state at the end of the very first cycle. This is expected since the majority of spins is isolated or form couples and such antiferromagnetic systems are known to have a RPM \cite{r14}. As $d$ increases the cycle opening $\Delta S$ becomes nonzero. However, a well pronounced increase of $\Delta S$ is observed only after the percolation threshold of the network at $d\approx 1$, when a giant spin cluster appears \cite{r9}. Then, after reaching a maximum, $\Delta S$ starts to decrease to zero as $d$ approaches the network size $N$ with the limit $\Delta S = 0$ obtained for fully connected network $d = N$. Such behavior is expected since the fully connected spin network can be viewed as a mean-field Preisach model, which has been shown to have RPM \cite{r15}.\\
\indent Numerical tests for different network sizes $N$ (up to $50^2$), different disorder $\sigma$ and different average energy per spin $\Delta=\bigl\langle J\sum_jA_{ij}s_j\bigr\rangle\approx J\cdot d\cdot\langle S\rangle$, revealed that the cycle opening $\Delta S$ depends on the ratio $\Delta/\sigma$ rather than on $\Delta$  and $\sigma$ separately. In addition, $\Delta S$ was observed to be independent of the network size as long as $d << N$, and we found no dependence on the mean $\mu$ in the assumed limit $\mu >>\sigma$. Since the probability of finding various topological interconnection structures in the ER network, such as trees and cliques (completely interconnected sub-graphs) \cite{r9}, depends both on $d$ and $N$, these results demonstrate that the first cycle opening does not depend on the topological properties of the network. We note, that this conclusion is supported by similar results obtained for a regular lattice binary spin models, where $\Delta S$ was calculated as a function of the interaction range (which is, of course, proportional to the coordination number) \cite{r16}.
\begin{figure}
\includegraphics[scale=0.33]{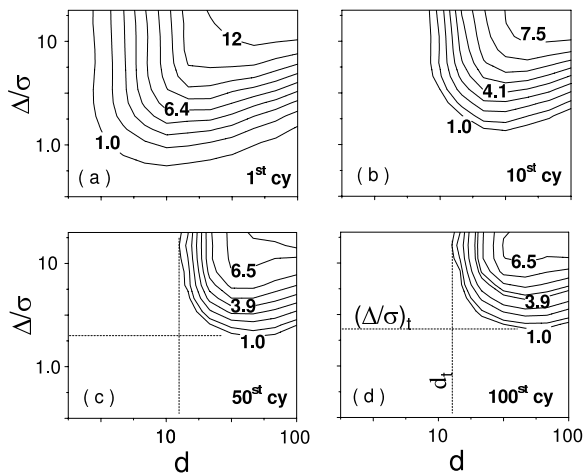}
\caption{\label{fig:epsart2} a)-d) contour maps showing the cycle opening $\Delta S$ for different values of interaction energy $\Delta/\sigma$ and the network degree $d$ (note the logarithmic scale of $\Delta/\sigma$ and $d$ axes) for respectively: 1-st, 10-th, 50-th and 100-th cycle. The lower bounds for the `limiting' region with non-converging loops correspond to $(\Delta/\sigma)_t\approx 2.3$ and $d_t\approx 13$. All minor loops have been obtained by reversing the field at $H_r$ where $\langle S\rangle = 0.2$, when the effect was most pronounced. Data are for $N = 50^2$, $\sigma = 0.1$ and averaged over 50 random graph and disorder realizations. Error bars level is about 1\%.}
\end{figure}\\
\indent It turns out, however, that the network topology does determine the minor loop formation behavior. To show this we studied the dependence of $\Delta S$ on the number of external field cycles, the network size $N$, the $\Delta/\sigma$ ratio and the connectivity parameter $d$. An example of the $\Delta S(\Delta/\sigma,d)$ function is given by the contour plots in the Fig. 2 a-d for four subsequent field cycle numbers and a fixed network size $N = 50^2$. The figure illustrates that in the low $(\Delta/\sigma,d)$-parameter region closed minor loops with $\Delta S=0$ appear already after a few initial field cycles (Fig.2 a-b). Similar behavior, termed as `tilting effect' (bascule), is frequently observed in magnetism \cite{r2} especially for clean ferromagnets. It is often attributed to the coupling between only a small numbers of magnetic domains (consistently with the low connectivity network picture considered here).\\
\indent For the parameter region bounded from below by certain critical values $(\Delta/\sigma)_t$ and $d_t$ (Fig.2 d) the behavior changes dramatically and closed minor loops often do not form even after $100$ field periods. Quite surprisingly, we find that the value $d_t$ is remarkably close to the theoretical value at which the ER network is known to undergo a topological transition associated with the emergence of cliques of size $4$ (fully interconnected groups of $4$ spins) \cite{r9}. Two new types of cyclic behaviors are observed: cycles with a very long transient time, $t_T > 100$, and subharmonic cycles with a spin pattern recovering repeatedly with period $t_S > 1$ ($t_T$ and $t_S$ are defined in the Fig. 1). Note that subharmonic cycles were also observed previously for some spin-glass networks \cite{r8}. Here, the period $t_S$ is found to grow with the number of 4-cliques present in the network. However, this growth is substantially slower than the increase of the network size $N$ and no conclusions can be made regarding the behavior of the sub-harmonic cycle length in the thermodynamic limit $(N\rightarrow\infty)$.\\
\indent On the other hand, the transient length grows rapidly as shown in the Fig. 3a by the log-log dependence of average $\langle t_T\rangle$ on the average number of 4-cliques, $\langle C_4\rangle$, (averaged over $100$ network realizations) for sufficiently strong interaction. Dependence on the network size is plotted in the Fig.3b and follows the exponential law $\langle t_T\rangle \propto \exp(N/\tau)$. The parameter $\tau$ has been obtained by fitting and, according to the inset in Fig.3c, $\tau\rightarrow\infty$ in the limit $\langle C_4\rangle\rightarrow 0$. This means that for $d > d_t$, when 4-cliques emerge, the $\langle t_T\rangle$ grows faster than $N$, implying that hysteretic trajectories will not converge in the thermodynamic limit $(N\rightarrow\infty)$.
\begin{figure}
\includegraphics[scale=0.73]{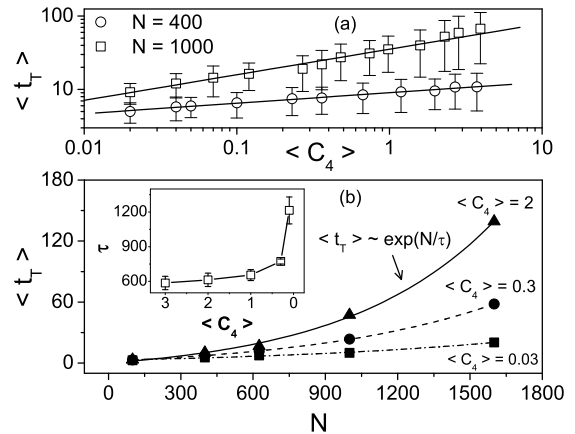}
\caption{\label{fig:epsart} a) Dependence of the transient length on the number of 4-cliques for $N = 20^2$ and $10^3$. $\langle t_T\rangle$ and $\langle C_4\rangle$ are averages over 100 network realizations. b) $\langle t_T\rangle$ as a function of $N$ obtained for $\langle C_4\rangle = 0.03$, $0.3$ and $2$. Lines for different $\langle C_4\rangle$ are exponential fits. Data correspond to reversal magnetization $\langle S\rangle = 0.2$, $\Delta/\sigma = 10$, and $\sigma = 0.1$. Inset: dependence of the exponential fit parameter $\tau$ on $\langle C_4\rangle$. $\tau$ grows without bound with diminishing number of 4-cliques $C_4$.}
\end{figure}\\
\indent It is surprising that the present antiferromagnetic binary spin network model with a simple single spin-flip field driven dynamics displays several qualitatively very different regimes, including a regime where hysteretic trajectories do not converge. Assuming that interactions are strong enough compared with the disorder, the determining factors appear to be the network connectivity and its topology. This suggests possible design rules for obtaining desired hysteretic behavior in artificial materials. Moreover, this conclusion suggests that behavior of hysteretic cycles could prove to be a useful characterization method for probing topology and connectivity of some systems with complex interactions. Care is required in carrying out such characterization because, due to noise, non-convergent cyclic behavior can be easily missed if only macroscopic measurements are performed. For example, average state (magnetization) in our simulations did not show the same behavior as the microscopic state. It is also interesting to note, that the model employed here did not reproduce some other forms of non-convergent behavior such as the reptation effect \cite{r2} observed in some magnetic systems. It is possible, that the reptation effects will be observable only in spin networks with mixed, i.e. ferromagnetic and antiferromagnetic interactions or in networks where spins have vector nature.\\
\indent We thank to Professors J.P. Sethna and H. Katzgraber for useful comments and suggestions. This work is supported in part by the NSF grant ECS-0304453.


\begin{thebibliography}{99}
\bibitem{r12}
	Strictly speaking, due to the truncation of the random numbers below $\alpha_{\min}$, the threshold distribution slightly differs from the Gaussian. However, the probability for observing a random number below $\alpha_{\min}=\mu-3\sigma$ is negligible.
\end{thebibliography}
\end{document}